\documentclass[useAMS,usegraphicx]{mn2e}
\usepackage{times}

\title[internal shocks in compact jets]{Internal shocks at the origin of the flat spectral energy distribution of compact jets}
\author[J. Malzac]{Julien Malzac$^{1,2}$\thanks{E-mail: julien.malzac@irap.omp.eu}\\
$^{1}$  Universit\'e de Toulouse; UPS-OMP; IRAP;  Toulouse, France\\
$^{2}$  CNRS; IRAP; 9 Av. colonel Roche, BP 44346, F-31028 Toulouse cedex 4, France}
\begin{document}

\date{Accepted 2009 October 10,  Received 2009 October 9; in original form 2009 April 2}

\pagerange{\pageref{firstpage}--\pageref{lastpage}} \pubyear{2008}

\maketitle

\label{firstpage}

\begin{abstract}

An internal shock model is proposed to interpret the radio to infrared (IR) emission of  the compact jets observed in the hard spectral state of X-ray binaries. Assuming that the specific bulk Lorentz factor of the jet at its base varies with a flicker noise power spectrum (i.e.  $P(f)\propto 1/f$), I estimate the energy dissipation profile along the jet and the resulting partially self-absorbed synchrotron emission.  For this type of velocity fluctuations, and a conical jet geometry, the shock dissipation at large distance from the black hole balances exactly the adiabatic losses. This leads to a flat radio to IR spectral energy distribution similar to that observed in compact jets. 
\end{abstract}


\section{Introduction}

Internal shocks caused by fluctuations of the outflow velocity are widely believed to power the multi-wavelength emission of jetted sources such as $\gamma$-ray bursts (Rees \& Meszaros 1994; Daigne \& Moscovitch 1998),  active galactic nuclei (Rees 1978; Spada et al. 2001), or microquasars (Kaiser, Sunyaev \& Spruit 2000; Jamil et al. 2010).
Internal shocks models usually assume that the jet can be discretised into homogeneous ejectas. Those ejectas are injected at the base of the jet with variable velocities and then propagate along the jet. At some point, the fastest fluctuations start  catching up and merging with slower ones. This leads to shocks in which a fraction of the bulk kinetic velocity of the shells is converted into internal energy. Part of the dissipated energy goes into  particles acceleration, leading to synchrotron and also, possibly, inverse Compton emission. 

Here, we will focus on the applicability of this model to the steady compact jets observed in the hard X-ray spectral state of black hole and neutron star binaries. 
These sources have an approximatively flat Spectral Energy Distribution (SED) extending from the radio to the mid-IR (e.g. Fender et al. 2000; Corbel \& Fender 2002; Chaty et al. 2003). These flat spectra are usually ascribed to self-absorbed synchrotron emission from conical compact jets (Blandford \& K\"onigl 1979) under the assumption of continuous energy replenishment of the adiabatic losses.  The compensation of these energy losses is crucial for maintaining this specific spectral shape (Kaiser 2006), and this can possibly be achieved through internal shocks.   Recently, however, Jamil et al. (2010) developed an internal shock model  for the emission of jets in X-ray binaries, and concluded that energy dissipation only through internal shocks is not enough to produce a flat SED.

 In fact this result is not very surprising. Let us  consider a jet with uniform time-averaged bulk Lorentz factor $\gamma=1/\sqrt{1-\beta^2}$ and a section of the jet consisting in a cylindric shell of radius $R$ and height $H\ll R$. This shell propagates downstream, so that at a time $\tilde{t}$, measured in the co-moving frame,  its position along the jet is $z=\gamma\beta c\tilde{t}$.
The function $R(\tilde{t})$ defines the jet geometry.
For simplicity, in this paper we will consider only conical jets: i.e. $R=R_b+z \tan{\phi}$, 
where $\phi$ is the half-opening angle of the jet, and $R_b$ is the radius of the jet at its base.
The shell has a specific internal energy $\tilde{\epsilon}$ (measured in the co-moving frame).
The energy losses are usually dominated by adiabatic expansion i.e. the pressure work exerted by the shell against the external medium as it travels downstream and expands:
 \begin{equation}
d\tilde{W}=Pd\tilde{V}= (\gamma_a-1) m\tilde{\epsilon}\frac{d\tilde{V}}{\tilde{V}}\simeq\frac{2m\tilde{\epsilon}}{3} \frac{dR}{R},
 \label{eq:alosses}
 \end{equation}
 where $P$ is the total pressure in the shell (including magnetic pressure), $\tilde{V}$ is its co-moving volume,  $m$ its mass and $\gamma_a$ is the effective adiabatic index of the flow. For simplicity, it is assumed here that all the components of the flow (thermal particles, relativistic particles, magnetic field...)  have the same adiabatic index corresponding to that  of a relativistic gas (i.e. $\gamma_a=4/3$). The last term of equation \ref{eq:alosses} is obtained neglecting the effects of possible longitudinal expansion. At large $z$, $R\simeq z\tan{\phi}$, and, in the comoving frame, the shell loses its specific energy at a rate: 
 \begin{equation}
 \left(\frac{d\tilde{\epsilon}}{d\tilde{t}}\right)_{ad}=-\frac{2(\gamma_a-1)\tilde{\epsilon}}{R} \frac{dR}{d\tilde{t}}\simeq-\frac{2\tilde{\epsilon}}{3\tilde{t}}.
\label{eq:alosses2}
\end{equation}
If, as required to reproduce the flat SED of compact jets, the adiabatic losses are fully replenished by some dissipative process, then $\tilde{\epsilon}$ must be a constant. Equations \ref{eq:alosses2} then implies that the replenishing mechanism must have a dissipation profile $\propto \tilde{t}^{-1}$ over a long section of the jet.  Usual internal shock models predict instead  a dissipation profile scaling like $\tilde{t}^{-5/3}$ (Beloborodov 2000) and therefore  cannot compensate for the losses over a long section of the jet.
 
Nevertheless, most studies of the internal shock model so far, including that of Beloborodov (2000) and Jamil et al. (2010), have implicitly assumed that the Fourier Power Spectral Density (PSD) of the velocity fluctuations injected at the base of the jet is flat (i.e. white noise).  In fact, the energy dissipation profile of the internal shocks is very sensitive to the shape of the PSD of the velocity fluctuations.  Indeed, let us consider a fluctuation of the jet velocity of amplitude $\Delta v$ occuring on a time scale $\Delta t$. This leads to the formation of a shock at a downstream distance  $z_s \propto \Delta t /\Delta v$. In this shock the fraction of the  kinetic energy  converted into internal energy will be larger for larger $\Delta v$. From these simple considerations we see that the distribution of the velocity fluctuation amplitudes over their time scales (i.e. the PSD)  is going to determine where and in which amount the energy of the internal shocks is deposited.  
In this paper, instead of a white noise PSD, we will assume that  the  PSD of  the injected fluctuations decreases with Fourier frequency like $P(f)\propto 1/f$ (i.e.  flicker noise). Such noise  occurs in many physical, biological and economic systems { (see e.g. Press 1978)}. In astronomy, it is observed in the solar activity {(e.g. Ryabov et al. 1997)} and, most notably, in the X-ray variability of X-ray binaries { (see e.g. Gilfanov 2010)}. In Section~\ref{sec:disprofile}, we show that shocks driven by this type of fluctuations can compensate exactly for the adiabatic losses in a conical geometry. In Section~\ref{sec:fc}, we estimate the jet volume filling factor. Then, in Section~\ref{sec:sed}, we derive estimates for  the predicted  SED of the jet. Finally, in Section~\ref{sec:application}, these results are discussed in the context of Black Hole Binaries (BHBs).

\section{Internal shock dissipation profile}\label{sec:disprofile}

 We assume that small, time dependent  fluctuations of the Lorentz factor are continuously injected at the base of the jet. The fluctuations depends on time $t$  and have a Fourrier power density spectrum $S(f)=S_0 f^{-1}$ for  frequencies ranging from $f_1$ to $f_0$. 
The initial variance of the injected fluctuations is therefore:
\begin{equation}
\gamma_{\rm rms0}^2=\int_{f_1}^{f_0}2S(f)df=2S_0\ln{(f_0/f_1)}.
\end{equation}
We model the jet as a set of discrete homogeneous shells that are ejected with such variable velocities at time intervals $\Delta t=(2f_0)^{-1}$. For simplicity, the mass of the ejectas is assumed to be a constant $m_0=0.5P_j\Delta t/(\gamma-1)c^2$, where $P_j$  is the total kinetic power of the two-sided jet.  At injection, the center of momentum of two neighbouring shells are separated by a distance $\lambda_0=\beta c \Delta t$. This length also corresponds to the smallest scale of the velocity fluctuations.  
As the ejectas propagate downstream, the fastest shells catch up and merge with slower ones. During this process of hierarchical merging the mass of the ejectas and their separation will increase $\lambda(t)/\lambda_0=m(t)/m_0=K(t)$. This growth in length scale implies a damping of the fluctuations of frequencies higher than $f_0/K$. As a consequence, the variance of the fluctuations decreases: 
\begin{equation}
\gamma_{\rm rms}^2(K)=\int_{f_1}^{\frac{f_0}{K}}2S(f)df=\gamma_{\rm rms0}^2-2S_0\ln{K}.
\label{eq:gammadek}
\end{equation}

The problem appears simpler when viewed in the frame moving with a Lorentz factor $\gamma$.  Throughout this paper, the tilded quantities are measured in this co-moving frame. In the limit of small scales fluctuations, this frame coincides with the frame of the centre of momentum of the shells. In this frame, the average velocity of the shells is 0 and its rms amplitude is, to first order in $\gamma_{\rm rms}/\gamma$,
\begin{equation}
\tilde{v}_{\rm rms}(K)=\frac{\gamma_{\rm rms}(K)c}{\gamma \beta}.
\label{eq:vrms}
\end{equation}
In the limit of low amplitude fluctuations, the velocities in the moving frame are non-relativistic. 
All of the kinetic energy of the shells is available for conversion into internal energy. 
We  define the specific free energy of the system (i.e. available for dissipation) as $\tilde{u}(K)=\tilde{v}^2_{\rm rms}(K)/{2}$.  Combining equations~\ref{eq:gammadek} and  \ref{eq:vrms} we see that the specific free energy decreases as the scale of the fluctuations increases:
\begin{equation}
\frac{d\tilde{u}}{dK}=-\frac{S_0 c^2}{K\gamma^2\beta^2}.
\label{eq:dusdk}
\end{equation}
{ The fluctuations of scale $K$ collide and merge after travelling during a time: 
\begin{equation}
\tilde{t}(K)=y\tilde{\lambda}/\Delta\tilde{v},
\label{eq:tdek}
\end{equation}
where $y$  is a factor of the order of unity accounting for  the effects of the longitudinal extension of the ejecta. It will be estimated in section~\ref{sec:fc}. $\Delta \tilde{v}$ is the average (absolute) velocity difference between the merging shells.} In the case of a linear variability process it can be estimated as:
\begin{equation}
{\Delta{\tilde{v}}}^2=  G^2(K)\frac{S_0c^2}{2\gamma^2\beta^2},
\label{eq:delv}
\end{equation}
with
\begin{equation}
G(K)^2=8\int_{f_1}^{f_0/K} \left[1-\cos{(K \pi f/f_0)}\right] f^{-1} df.
\end{equation}
In the limit  $K \ll f_0/f_1$, the function $G(K)$ is nearly constant: 
\begin{equation}
G(K)\simeq \mathcal{M}=\left[{-8\sum_{n=1}^{+\infty}\frac{(-\pi^2)^n}{(2n)! (2n)}}\right]^{\frac{1}{2}}\simeq 3.6
\label{eq:m}
\label{eq:g0}
\end{equation}
Then, combining equations~\ref{eq:tdek}, \ref{eq:delv}, \ref{eq:g0}, we find that
the growth of the fluctuations is linear: $K=\tilde{t}/\tilde{t_0}$,
with a characteristic time-scale:
\begin{equation} 
\tilde{t_0}=\sqrt{2}y\gamma^2\beta^2 f_0^{-1}\mathcal{M}^{-1}S_0^{-\frac{1}{2}}
\end{equation}
Then combing this with equation~\ref{eq:dusdk}, we conclude that 
at times $\tilde{t_0}\ll \tilde{t} \ll \tilde{t_0}f_0/f_1$, the free energy is dissipated at a rate:
\begin{equation}
\frac{d\tilde{u}}{d\tilde{t}}=-\frac{S_0 c^2}{\gamma^2\beta^2 \tilde{t}}.
\end{equation}
This scaling of the dissipation profile balances the adiabatic losses (equation~\ref{eq:alosses2}), for:
\begin{equation}
\tilde{\epsilon}=\tilde{\epsilon}_s=\frac{1}{(\gamma_a-1)}\frac{S_0 c^2}{2\gamma^2 \beta^2}.
\label{eq:epsilon}
\end{equation}
From equation~\ref{eq:epsilon}, we can estimate the sound speed in the jet as $\tilde{v}_{\rm s}=\sqrt{\tilde{\epsilon}(\gamma_a-1)}$. Note that, since $\tilde{\epsilon}$ also contains the magnetic energy,  $\tilde{v}_{\rm s}$  represents in fact an upper limit on the true sound velocity.  Using equation~\ref{eq:delv} and~\ref{eq:m},  we find the average Mach number of the colliding ejectas is at least $\Delta \tilde{v}/v_{s}=\mathcal{M} \simeq 3.6$. Such supersonic collisions, ensure that strong shock waves will be generated. The region of the jet with significant shock dissipation starts at a distance $z_0\simeq\gamma\beta c \tilde{t_0}$. Therefore, at  $z<z0$,  $\epsilon\simeq0$.
At the other end of the jet, at distances $z>z_f=z_0f_1/f_0$, all the energy of the internal shocks has been dissipated, and the flow simply cools down through adiabatic losses: $\tilde{\epsilon}\sim \tilde{\epsilon}_{s} (z/z_f)^{-2/3}$.
In the range $z_0$--$z_f$, the specific energy of the flow  is  approximately uniform and is given by equation~\ref{eq:epsilon}.

\begin{figure*}
 \includegraphics[width=0.97\linewidth]{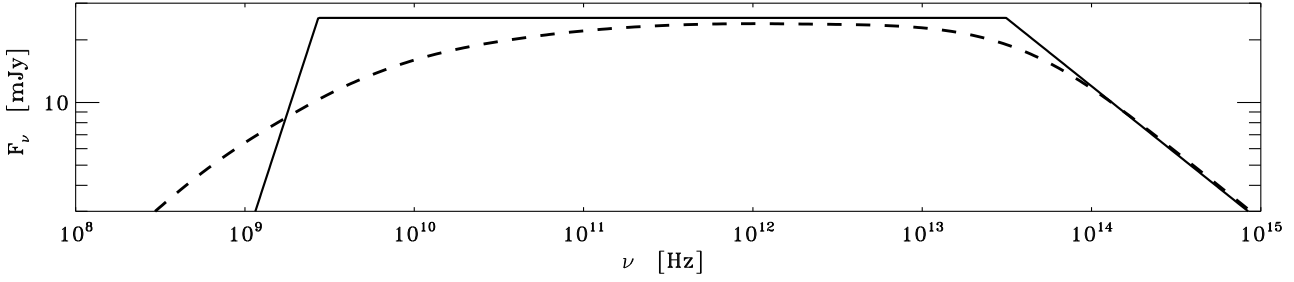}
 \caption{Typical SED obtained for a source at a distance of 2 kpc and an inclination of 40$\degr$. The dashed curve shows the result of a Monte-Carlo simulation, the full lines shows the analytical approximation. The jet lorentz factor is $\gamma=2$ for a jet kinetic power $P_J=10^{-2} L_E$  fluctuating with a fractional rms amplitude of 30 percent with a flicker noise PSD at Fourrier frequencies ranging from $f_1=1.6 \times 10^{-3}$ Hz to $f_0=50$ Hz.  The jet radius at the base is $R_b=10 R_G$ and its half opening angle is $\phi=1\degr$.  The equipartition ratio is $\xi=1$, the electrons in the jet have a power-law energy distribution with slope $p=2.3$ between $\gamma_{\rm min}=1$ and  $\gamma_{\rm max}=10^{6}$. The emission from the counter jet is not shown.}
 \end{figure*}

\section{Longitudinal extension of the ejectas}\label{sec:fc}

The volume filling factor of the colliding shells, $f_{\rm v}$, is defined so that the average length of a shell of separation $\lambda$ is $f_{\rm v}\lambda$.
If $f_{\rm v}\ll1$, the jet is constituted of thin colliding `pancakes', while for $f_{\rm v}=1$, the jet is a continuous flow.  In principle, $f_{\rm v}$ depends on the properties of the shells at injection. But $f_{\rm v}$ is also expected to depend on $z$ (or $\tilde{t}$). Indeed, as the shells propagate and gain internal energy  through shocks, they also expand under the effect of their own pressure. Here, as it is customary, we assume that this expansion occurs at the sound speed in the co-moving frame $\tilde{v}_{\rm s}$.  For simplicity, we will  consider that the shells  expand in near vacuum,  and therefore the adiabatic losses due to longitudinal expansion can be neglected (free expansion). Any pressure work done during this expansion can only be used to accelerate or compress neighbouring shells so that overall, this does not represent  a loss of energy for the flow. We thus do not expect that taking these effects into account would change the results dramatically.
 
 The shells can also undergo significant compression during shocks. When two shells are in the process of merging  the compression velocity is simply the difference of velocity of the two merging ejectas. During a collision however, the compression lasts only for the time necessary for the shock to cross the shell, that is $\simeq x f_{\rm v}\tilde{\lambda}/\Delta \tilde{v}$, where $x^{-1}\Delta \tilde{v}$ is the speed of the shock in the frame of the unshocked ejecta. Applying the jump conditions of Blandford \& McKee (1976) and to first order in $\Delta \tilde{v}$, their equation (5) gives:
 \begin{equation}
 x^{-1} \simeq \sqrt{1/4+\gamma_a/2+(\gamma_a/2)^2}=7/6.
 \end{equation}
Since  the collisional time-scale is  $\sim y\tilde{\lambda}/\Delta\tilde{v}$, the shells are compressed during a fraction~$\simeq xf_{\rm v}/y$ of the time, and the time-averaged compression velocity is reduced to $\tilde{v}_{\rm comp}\simeq xf_{\rm v}\Delta\tilde{v}/y$. The parameter $y$ in equation~\ref{eq:tdek} can be approximated as: 
\begin{equation}
y=\frac{1-f_{\rm v}}{1+2\mathcal{M}^{-1}}.
\label{eq:y}
\end{equation}
The numerator on the right hand side of equation~\ref{eq:y} accounts for the reduced distance between shells due to their longitudinal extension. The denominator corrects for the increased collision rate caused by the expansion velocity of the shells.
  
As shown in Section~\ref{sec:disprofile},  the choice of a flicker noise, implies that, at  large $z$, the expansion velocity is a constant and independent of $f_{\rm v}$. The compression velocity instead increases monotonically with $f_{\rm v}$. As a consequence the volume filling factor evolves toward an equilibrium in which, on average,  compression balances longitudinal expansion. This occurs for 
\begin{equation}
f_{\rm v}\simeq[1+x(1+\mathcal{M}/2)]^{-1}\simeq 0.3,
\end{equation}
independently of the parameters. In fact, the evolution of $f_{\rm v}$ is more complex, but in practice, assuming it is a constant turns out to be a good approximation. We will therefore set $f_{\rm v}=0.3$ and $y=0.5$.

\section{Spectral energy distribution}\label{sec:sed}
{ We now use a simple emission model to estimate the properties of the jet SED as a function of the amplitude and bandwidth of the Lorentz factor fluctuations.}
A fraction of the internal energy is in the form of relativistic electrons.
For simplicity we assume that the distribution of the electrons on their Lorentz factors $\gamma_e$,  is a power law:  $n(\gamma_e)=n_0\gamma_e^{-p}$ with $\gamma_e$ in the range $\gamma_{\rm min}$-- $\gamma_{\rm max}$.
The remaining fraction is in the form of magnetic field energy. The particle energy density is:
\begin{equation}
n_0 m_ec^2\int_{\gamma_{\rm min}}^{\gamma_{\rm max}}\gamma_e^{1-p}d\gamma_e=n_0 m_ec^2/i_\gamma=\xi B^2/8\pi,
\end{equation}
where $\xi$ is the equipartition factor that we assume constant. The magnetic field energy density is:
\begin{equation}
\frac{B^2}{8\pi}=\frac{\tilde{\rho}\tilde{\epsilon}}{1+\xi},
\end{equation}
where the mass density along the jet  is given by:
\begin{equation}
\tilde{\rho}=\frac{P_{J} R^{-2}}{2f_{\rm v}(\gamma-1)c^2 \gamma\beta c \pi }.
\end{equation}
As a consequence, in this regime, the magnetic field scales as: $B=B_0 R_0/R$, where $R_0=z_0\eta\tan \phi$, and:
\begin{equation}
B_0=\frac{2f_0S_0\mathcal{M}} {y \eta \tan{\phi}}\sqrt{\frac{P_J(\gamma\beta)^{-9}(\gamma-1)^{-1}}{ (\gamma_a-1)(1+\xi)f_{\rm v} c^3}},
\label{eq:b0}
\end{equation}
where $\eta=1+R_b/(z_0\tan \phi)\sim 1$.
This scaling holds approximately for $r\gg R/R_0 \gg 1$, where: $r=f_0/f_1+1-\eta^{-1}\simeq f_0/f_1$.
At larger distances, i.e. $ R> r R_0$, the magnetic field decays faster due to the absence of shock dissipation to balance the adiabatic losses. 

The relativistic electrons emit  synchrotron radiation. Their emissivity at a co-moving photon frequency $\tilde{\nu}$ is (Rybicki \& Lightman 1979):
\begin{equation}
j_{\tilde{\nu}}=K_j\xi B^{\frac{p+5}{2}}\tilde{\nu}^{-\frac{p-1}{2}},
\end{equation}
where
\begin{equation}
K_j=\frac{\sqrt{3}e^{3}i_{\gamma} \left(\frac{m_ec}{3 e}\right)^{-\frac{p-1}{2}}}{16\pi^2 m_e^2 c^4 (p+1)} \Gamma\left(\frac{3p+19}{12}\right)\Gamma\left(\frac{3p-1}{12}\right),
\end{equation}
and  $\Gamma$ is the usual gamma function. This emission can be self-absorbed. The absorption coefficient is :
\begin{equation}
\alpha_{\tilde{\nu}}=K_{\alpha}\xi B^{\frac{p}{2}+3}\tilde{\nu}^{-(p+4)/2},
\end{equation}
\begin{equation}
K_\alpha=\frac{\sqrt{3}e^3 i_{\gamma}}{64\pi^2 m_e^3 c^4}\left(\frac{3e}{2\pi m_e c}\right)^{p/2}\Gamma\left(\frac{3p+2}{12}\right)\Gamma\left(\frac{3p+22}{12}\right).
\end{equation}
The specific intensity at the surface of the jet is:
\begin{equation}
I_{\tilde{\nu}}=\frac{j_{\tilde{\nu}}}{\alpha_{\tilde{\nu}}}\left(1-e^{-\alpha_{\tilde{\nu}}R}\right).
\label{eq:inu}
\end{equation}
In the case of a conical jet in a steady state, making an angle $\theta$ with the line of sight, the flux at an observed frequency $\nu=\delta\tilde{\nu}$, is:
\begin{equation}
F_{\nu}(\nu)=\frac{f_{\rm v} \delta^2}{2D^2\tan{\phi}}\int_{R_b}^{+\infty}R I_{\tilde{\nu}}(\tilde{\nu})dR,
\end{equation}
where $\delta=\left[\gamma(1-\beta\cos\theta)\right]^{-1}$ is the Doppler factor and $D$ the distance to the source. 
In order to obtain analytical estimates of the observed SED we now assume that the magnetic field has the $R^{-1}$ dependence given by equation~\ref{eq:b0} for $R$ between  $R_0$ and $R_f=r R_0$, and $B=0$ elsewhere.
The SED can then be approximated as follows: 
At the highest frequencies the emission is optically thin and the spectrum is a power law $F_{\nu}\propto \nu^{-\frac{p-1}{2}}$, while at the lowest frequencies the emission is optically thick $F_{\nu}\propto \nu^{\frac{5}{2}}$.
For $p>-3$, there is a range of intermediate frequencies, comprised between $\nu_s$ and $\nu_t$, in which the emission at a given frequency is a mixture of optically thick and thin emission from different regions of the jets. In this regime,  the flux is independent of the photon frequency (see e.g. Kaiser 2006) :
\begin{equation}
F_{\nu}(\nu)\simeq\frac{-\Gamma(a)\delta^{2}f_{\rm v}K_j \left(R_0B_0\right)^{2-a}}{D^2\tan{\phi}(p+4)K_\alpha^{1+a}\xi^a},
\end{equation}
where $a=-5/(p+4)$.
The observed turnover frequency, $\nu_t$, is at the intersection of the optically thin and flat  asymptotic branches:
\begin{equation}
\nu_t\simeq\delta\left(K_{\alpha}\xi R_0 B_0^{\frac{p}{2}+3}\right)^{\frac{2}{p+4}}\left[\frac{r^{\frac{1-p}{2}}-1}{(a+1)\Gamma(a)}\right]^\frac{2}{p-1}.\label{eq:tnfreq1}
\end{equation}
The observed frequency of transition to the low frequency optically thick regime is given by: 
\begin{equation}
\nu_s\simeq\delta\left(K_{\alpha}\xi R_0 B_0^{\frac{p}{2}+3}\right)^{\frac{2}{p+4}}\left[\frac{r^{\frac{5}{2}}-1}{a\Gamma(a)}\right]^{-\frac{2}{5}}.
\label{eq:tnfreq2}
\end{equation}
These analytical estimates have been tested against Monte-Carlo simulations of the internal shock model with a code similar to that of Jamil et al. (2010). Fig. 1 compares the average SED given by the analytical formulae to that  obtained from a simulation with parameters corresponding to Cyg~X-1 (see caption and Section~\ref{sec:application}). The main differences occur around and below $\nu_s$ when the self-similar approximation that we used in the analytical formulae breaks down. Also, unlike the simulation, the analytical model ignores the emission of from the jet at distance larger than $z_f$. 

{ It is worth noting, however, that in order to obtain analytical estimates of the SED, we had to make several simplifications. First, equation~\ref{eq:inu} neglects the angular dependence of the optical depth, this is a good approximation only for systems with a large inclination. Then, we also assumed that the electron energy distribution is a pure power-law and therefore neglected the effects of a thermal component in the electron distribution. We also neglected the effects of adiabatic cooling on the shape of the electron distribution. Simple estimates indicate that both effects will affect the SED only marginally. On the other hand the effects of radiative cooling can be very strong in the region close to the base of the emitting region (see e.g. Chaty et al. 2011). These radiation losses, that we have neglected, affect predominantly the most energetics electrons, emitting in the optically thin regime, while the flat optically thick part of the SED  is essentially unchanged (Zdziarski et al. 2012).}

\section{Application to black hole binaries}\label{sec:application}
 
We can now use the formalism developed in the previous section to estimate the the jet parameters and emission properties in BHBs. 
In the following we assume a black hole of mass $M=10 m_{1}  M_{\odot}$ and the radius of the jet at $z=0$ is: 
\begin{equation}
R_b=10 r_1  R_G= 1.5 \times 10^{7}   r_1 m_{1} \quad {\rm  cm}.
\end{equation}
If the jet is launched by the accretion disc, the highest and lowest frequencies of variability $f_0$ and $f_1$ must relate to time scales of the accretion flow.  The picture could be similar to that of the model by Lyubarski (1995) in which density fluctuations generated at a range of disc radii propagate inward and lead to flicker noise fluctuations of the emission of the inner part of the accretion flow.  Similar fluctuations may also lead to a variable velocity of the ejected shells. The detail of how the disc fluctuations are transferred to the jet is out of the scope of  this paper. {  An interesting possibility is that the rapid episodic ejections could be produced through MHD mechanisms similar to that leading to coronal mass ejections in the sun (Yuan et al. 2009).  Let  $g$ be the ratio of the ejection time scale $\Delta t=f_0^{-1}/2$  to the Keplerian time-scale of the accretion disc at radius $R_b$. In our numerical estimates we will assume that those time-scales are comparable (i.e. $g\sim1$):}
 \begin{equation}
f_0\simeq  50 \quad  r_1^{-3/2}  m_{1}^{-1} g^{-1} \quad {\rm Hz}.
\end{equation}
Similarly we associate the lowest frequency of the fluctuations to $2g$ times the dynamical time-scale at a large, possibly outer, radius of  the disc   $R_d$:
\begin{equation}
R_d=10 ^{5} r_5  R_G=1.5 \times 10^{11} r_5 m_{1} \quad {\rm cm},
\end{equation}
\begin{equation}
f_1\simeq  5 \times 10^{-5} \quad  r_5^{-3/2}  m_{1}^{-1} g^{-1} \quad {\rm Hz}.\label{eq:f1}
\end{equation}
We will assume that  the fractional rms amplitude of  fluctuations of the kinetic power is $\sigma=\gamma_{\rm rms0} /(\gamma-1)\sim 0.3$,  i.e.  comparable to that observed in the X-ray variability of black hole binaries in the hard state. The average travel time of a shell from the ejection point to the first shocks is then:  
\begin{equation}
\tilde{t}_0\simeq   90 \quad \frac{ \gamma+1}{3}  \frac{ g r_1^{3/2} m_{1} \zeta^{1/2}} {\sigma_{0.3}} \quad{\rm ms}, 
\end{equation}
where  $\sigma_{0.3}=\sigma/0.3$, and  $\zeta=\ln{(R_d/R_b)}/\ln{10^4}$. { This time-scale might be related to the delays of comparable amplitude  observed between the  X-ray emission of the accretion flow and the optical and IR  emission from the jet in GX339-4  (Casella et al. 2010; Gandhi et al. 2010) or XTE J1118+480 (Kanbach et al. 2001; Hynes et al. 2003; Malzac et al. 2003)}.  It corresponds to a lowest distance at which significant emission takes place of:
\begin{equation}
z_0 \simeq 5 \times 10^9 \quad \frac{ (\gamma+1)\gamma\beta}{\sqrt{27}}  \frac{ g r_1^{3/2} m_{1} \zeta^{1/2}} {\sigma_{0.3}} \quad{\rm cm}. 
\end{equation}
This is comparable to the distance of $\sim 10^3 R_g$ inferred  in Cyg~X-1 both by the modelling the of SED (Zdziarski et al. 2012), and that of the orbital modulation of the radio emission (Zdziarski 2012). The size of the dissipation region is then 
\begin{equation}
z_f \simeq   5  \times 10^{15} \quad \frac{ (\gamma+1)\gamma\beta}{\sqrt{27}} \frac{ g r_5^{3/2} m_{1} \zeta^{1/2}} {\sigma_{0.3}}  \quad{\rm cm}, 
\end{equation}
which is roughly  comparable to the extension of the radio jet of Cyg~X-1 in the VLBA images of Stirling et al. (2001).  These authors also constrain the jet opening angle of Cyg~X-1 to be small $\phi=\phi_1 \frac{\pi}{180} <2\degr$. From this,  we can estimate the radius of the jet  at the base of the emitting region:  
 \begin{equation}
 R_0 \simeq  8 \times 10^7 \quad \frac{ (\gamma+1)\gamma\beta}{\sqrt{27}}   \frac{  \phi_1\eta g r_1^{3/2} m_{1} \zeta^{1/2}} {\sigma_{0.3}}  \quad{\rm cm}. 
\end{equation}
H${\alpha}$ and \mbox{[O\,{\sc iii}]} measurements of the optical nebula surrounding Cyg~X- 1  (Gallo et al.   2006 ; Russell et al. 2007) indicate that the  jet power is of the oder of a few percent of the Eddington luminosity $L_E$. This implies a magnetic field at the base of the emitting region: 
\begin{equation}
B_0 \simeq 2 \times 10^4 \quad \frac{\sigma_{0.3}^2 }{\phi_1\eta g \zeta}  \left[\frac{162\sqrt{3}}{1+\xi} \frac{(\gamma+1)^{-3}}{\gamma^3\beta^3}\frac{P_{\%}}{r_1^{3} m_{1}}\right]^{1/2} {\rm G},
\end{equation}
where $P_{\%}=\frac{P_J}{0.01 L_E}$. These  estimates for $R_0$ and $B_0$ are comparable to the values inferred in Cyg~X-1 (Zdziarski et al. 2012), GX 339-4 (Gandhi et al. 2011) and  XTE J1550-564 (Chaty et al. 2011). Fixing the parameters of the electron distribution to standard values ($p=2.3$, $\gamma_{\rm min}=1$, $\gamma_{\rm max}=10^6$) , we can now evaluate the flux of the flat section of the spectrum:
\begin{equation}
F_{\nu0}\simeq 30  \quad \frac{\delta^2 \xi^{\frac{5}{p+4}}}{D^2_{\rm kpc} \phi_1}\left[\frac{2\sqrt{27}}{1+\xi}\frac{   P_{\%} m_{1} \sigma_{0.3}^{2}}{ \zeta (\gamma+1) \gamma\beta}\right]^{\frac{2p+13}{2p+8}}
{\rm mJy},
\end{equation}
where $D_{\rm kpc}$ is the distance to the source expressed in kpc.
This estimate is consistent with the flux observed in bright hard states (e.g. at $\simeq$ 2 kpc, Cyg~X-1 has an average radio flux of of 15 mJ at 15~GHz). Moreover we see that for $p\simeq 2$  the flux depends on the jet power approximately as $F_{\nu0}\propto P_J^{1.4}$, as inferred from the observations of the radio-X-ray correlation (Gallo, Fender \& Pooley 2003). The turnover frequency is expected in the mid-infrared, in agreement with the observations of Cyg~X-1 (Rahoui et al. 2011) or GX 339-4 (Gandhi et al. 2011):
\begin{equation}
\nu_t\simeq 1.8  \times 10^{13} \quad  n_1 r_{1}^{-\frac{3}{2}} \quad {\rm Hz},
\end{equation}
where
\begin{equation}
 n_1=\frac{\delta \xi^{\frac{2}{p+4}}\sigma_{0.3}^{\frac{2p+10}{p+4}}}{\eta g \phi_1 m_{1}^{\frac{p+2}{2p+8}} \zeta^{\frac{p+5}{p+4}}}\left(\frac{2P_{\%}}{1+\xi}\right)^{\frac{p+6}{2p+8}}\left[\frac{3\sqrt{3}}{(\gamma+1)\gamma\beta}\right]^{\frac{3p+14}{2p+8}}.
 \end{equation} 
 Finally the model predicts that the emission should decrease significantly at frequencies below:
\begin{equation}
\nu_s\simeq 40  \quad  n_1 r_{5}^{-\frac{3}{2}} \quad  {\rm MHz},
\end{equation}
and this is something that can be investigated in the near future with LOFAR. Note that in Fig.~1,  $\nu_s$ occurs at much higher frequencies (a few GHz).  Indeed, because simulating  very long time-scale fluctuations is time consuming,  $f_1$ was set  to $1.6\times10^{-3}$ Hz in Fig.~1, rather than $\sim 5\times10^{-5}$ Hz as given by equation~\ref{eq:f1}.

We conclude that internal shocks driven by flicker noise fluctuations can produce not only the flat SED of  compact jets in BHBs, but also other properties such as the flux amplitude or the location of the break frequency. The model also predicts  strong multiwavelength variability that will be the focus of future works.
\section*{Acknowledgments}
The author thanks the Institute of Astronomy in Cambridge (UK) for hospitality. This work has received fundings from PNHE in France. The author is grateful to Poshak Gandhi, Andrzej Zdziarski and the anonymous referee for useful comments.


\label{lastpage}

\end{document}